\documentstyle[preprint,aps]{revtex}
\begin{document}
\draft
\title{ Fluctuations and scaling of inverse participation ratios \\
in random binary resonant composites }
\author{Y. Gu$^{1,2}$ and K. W. Yu$^2$}
\address{$^1$ State Key Laboratory for Mesoscopic Physics, Department of\\
Physics, \\
Peking University, Beijing 100871, China }
\address{$^2$ Department of Physics, The Chinese University of Hong Kong, \\
Shatin, New Territories, Hong Kong, China }
\author{Z. R. Yang}
\address{Department of Physics, Beijing Normal University, Beijing 100875,\\
China}
\maketitle

\begin{abstract}
We study the statistics of local field distribution solved by the
Green's-function formalism (GFF) [Y. Gu et al., Phys. Rev. B {\bf 59} 12847
(1999)] in the disordered binary resonant composites. For a percolating
network, the inverse participation ratios (IPR) with $q=2$ are illustrated,
as well as the typical local field distributions of localized and extended
states. Numerical calculations indicate that for a definite fraction $p $
the distribution function of IPR $P_q$ has a scale invariant form. It is
also shown the scaling behavior of the ensemble averaged $\left\langle
P_{q}\right\rangle $ described by the fractal dimension $D_q$. To relate the
eigenvectors correlations to resonance level statistics, the axial symmetry
between $D_2$ and the spectral compressibility $\chi$ is obtained.
\end{abstract}


\vskip5mm

The statistics of eigenfunctions of the random matrix with orthogonal,
unitary and symplectic symmetry has been well reviewed by Guhr$^{1}$ and
Mirlin$^{2}$. A strong fluctuation of eigenfunctions at the critical point
is one of the prominent hallmarks of Anderson metal-insulator transition.
These fluctuations can be represented by a set of inverse participation
ratios (IPR) $P_{q}=\int d^{d}r|\Psi (r)|^{2q}$. Wegner put forward from the
renormalization group treatment of the supersymmetry $\delta $ model in $%
2+\epsilon $ dimensions that IPR show the multifractal scaling$^{3,4}$ with
respect to the system size $L$, $P_{q}=AL^{-D_{q}(q-1)}$, in a good metal,
while in an insulator, $P_{q}\propto L^{0}$. For the critical power-law
random bounded matrix (PRBM) ensemble$^{5}$ with $\beta =1$, not only the
distribution function of the critical $P_{q}$ was found to have a scale
invariant form, but also the scaling exponent $D_{q}$ was obtained from the
shift of the distribution of IPR with the size $N$. At the criticality, the
multifractality of the ensemble-averaged $\ \left\langle P_{q}\right\rangle $
has been derived analytically in the limits of weak and strong couplings, as
well as calculated numerically in the full range of couplings$^{6}$.
Recently, local field distribution of resonant composites has attracted
great interest. Various optical eigenmode localizations, such as
surface-plasmon modes in the colloid clusters$^{7}$, localized dipolar
excitations on the roughly nanostructured surfaces$^{8}$, and selective
photomodification in the fractal aggregates of colloidal particles$^{9}$,
were reported to be relevant to the specific filed distribution. It was
found that the local field at the percolation threshold fluctuates from one
optical eigenmode to another$^{10,11}$. Some attempts to investigate the
scaling of the successive local electric field moments have been made$%
^{11-13}$. However, it is difficult to take into account the structure
sensitivity to local field distribution when resonance happens. In order to
study the statistics of resonant composites, one must answer the following
questions. How do the optical eigenmodes distribute in the resonant area?
For each mode, how about its local field distribution? And how to understand
the fluctuations and scaling of the local field distribution? It is known
that the resonance spectrum and local field distribution for each eigenmode
can be solved analytically using the Green's-function formalism (GFF)$%
^{14,15}$. The first question has been answered in view of level spacing
statistics$^{14,16}$. In this work, local field distributions of typical
localized and extended eigenstates are illustrated, and to answer the third
question, the statistics of local field distribution is numerically studied
from the random matrix theory (RMT).

In this work, a binary network is considered where the impurity bonds with
admittance $\epsilon _{1}$ are employed to replace the bonds in an otherwise
homogeneous network of identical admittance $\epsilon _{2}$. The admittance
of each bond is generally complex and frequency-dependent. All the impurity
bonds construct the clusters subspace. Using the GFF, the eigenvalues and
eigenvectors of Green's-matrix $M$ can be solved analytically$^{14,15}$.
Because $M$ maps the geometric configuration of the clusters subspace, it is
a random matrix where RMT can set in. Its solutions summarize all the
geometric resonances of the clusters subject to the external sources and in
the quasistatic limit. The element of $M$ is defined as $M_{{\bf x},{\bf y}%
}=\sum_{{\bf z}\in C({\bf y})}(G_{{\bf x},{\bf y}}-G_{{\bf x},{\bf z}})$ in
which ${\bf z}\in C({\bf y})$ means that the jointing points ${\bf z}$ and $%
{\bf y}$ belong to the clusters subspace and are the nearest neighbors, and $%
G_{{\bf x},{\bf y}}$ is the Green's function of Laplace operator on the
infinite square, i.e., $-\Delta G_{{\bf x},{\bf y}}=\delta _{{\bf x},{\bf y}%
} $ with $G_{{\bf x},{\bf x}}=0$. More clearly, $M_{{\bf x},{\bf y}}$
describes the interaction between ${\bf x}$ and ${\bf y}$, and is closely
related to the ``environment'' or the nearest neighbors of ${\bf y}$.
Different from the quantum systems, local field distribution is not directly
equal to the eigenvector of $M$. When resonance happens, the local field $V_{%
{\bf x}}$ can be expressed as a separate product of one sum of right
eigenvectors of $M$ and another sum of left eigenvectors that depends on the
source term. Motivated by that local field distribution may have the general
statistical features, this paper mainly devotes to the fluctuations and
scaling of the IPR.

In the following, we start from solving Green's-matrix $M$ numerically in
the disordered binary networks. The distribution functions of the IPR are
calculated for system size ranging from $L=16$ to $L=38$ and for various
values of $p$. For the ensemble averaging over $1000$ samples are used in
the percolating case at $L=30$, for each of which there are more than $750$
nontrivial eigenstates to be produced. In the calculation of the scaling
behavior of the IPR, $\left\langle P_{q}\right\rangle $ is averaged over $%
7\times 10^{5}$ right eigenvectors. The difference between using $1000$
samples of $30\times 30$ system and using a large system, say $300\times 300$%
, is that there is a minor correction in the calculations. Numerical
calculations indicate that the distribution function of IPR $P_{q}$ has a
scale invariant form for a definite $p$ within the interval $[0.3,0.5]$. It
is also found the scaling behavior of the ensemble averaged $\left\langle
P_{q}\right\rangle $ described by the fractal dimension $D_{q}$. The
multifractal scaling of $\left\langle P_{q}\right\rangle $ with respect to $%
L $ is obtained for $p=0.5$ and $p=0.4,$as well as the fractal behavior of $%
\left\langle P_{q}\right\rangle $ for $p=0.3$. Note that here $L$ represents
the scale of square network, while in Refs. \cite{5} and \cite{6}, $N$ is
the size of $N\times N$ matrix. To relate the level statistics to
eigenvectors correlations, the symmetry between $D_{2}$ and the spectral
compressibility $\chi $ is obtained.

In order to have a direct observation, figure 1 displays the values of IPR
of right eigenvectors with $q=2$ for a $40\times 40$ percolating network.
The peaks represent the localized states and valleys the extended states. It
is shown that peaks and valleys randomly distribute in the resonant area.
The localized states incline to accumulate near $s\rightarrow 0$ and $%
s\rightarrow 1.0$, while at about $s\rightarrow 0.5$, more extended states
are found. Various optical excitations$^{7-9}$, optical nonlinear
enhancements$^{13,17}$, and eigenmode localizations$^{18}$ are typical
localized states with the high values of IPR. For a localized state, it is
found that the value of $P_{2}$ of right eigenvector is always high, as well
as the IPR of potentials and electric fields$^{14}$. Therefore, it is
reasonable to represent the localization or extension of eigenstates only by
the IPR of right eigenvectors. When resonance happens, the inhomogeneous
local field around the impurity metallic clusters leads to a large
enhancement in the effective linear and nonlinear optical responses. It is
expected that the high values of IPR, or the localized states, correspond to
the strong nonlinear optical enhancements. There are an anomalous absorption
in the infrared area in percolating metal-dielectric thin films$^{11,20}$
and a large optical nonlinear enhancement in the region of high frequencies$%
^{14}$. In ref. \cite{14}, the separation of the absorption peak from
nonlinear enhancement peak is discussed. Different from the causes in the
dilute anisotropic networks$^{19}$, it originates in the self-similar
percolating structure.

Then, corresponding to the different values of IPR in Fig. 1, local field
distributions of typical localized and extended states are illustrated in
the forms of 3D plots. Fig. 2(a) shows a very localized state with the high
value of $P_{2}$. It is seen that when one eigenmode is excited, the rest of
the volume remains almost unexcited$^{7}$. For simplicity, those strong
localized fields are called ``hot spots''. In this figure, the brightest
``hot spot'' is found. In Fig. 2(b), an extended state with the low value of 
$P_{2}$ is plotted, where local fields are not uniform and many smaller and
weaker ``hot spots'' are found. In Ref. [14], we also shown an intermediate
state with the average value of $P_{2}$, where we can see more ``hot spots''
and local fields are very inhomogeneous. ``Hot spots'' are found to be
sensitive to the admittance ratio $h(=\epsilon _{1}/\epsilon _{2})$ and
external fields. They also lead to the photomodification$^{9}$ and huge
enhanced nonlinear responses$^{13,17}$.

When we plot the IPR of several samples with different size, it is intuitive
that the general feature of the distribution function of the IPR should
exist. Figure 3 displays the distributions of $\ln (P_{q}/\left\langle
P_{q}\right\rangle )$ for $p=p_{c}=0.5$, where $P_{q}$ is normalized by $%
\left\langle P_{q}\right\rangle $ and $q$ is ranged from $2$ to $8$. It is
seen that the shape of the distribution function changes with increasing $q$%
. It is also shown that for the distribution function of $\ln
(P_{q}/\left\langle P_{q}\right\rangle )$ the change of $p$ leads to the
shift and deformation of the shape$^{14}$. This is different from the recent
argument of Mirlin in which different parameter $b$ leads to the changes of
the distribution function of $\ln (P_{q})$ for the critical PRBM ensemble$%
^{6}$.

Figure 4 displays the distributions of $\ln (P_{2})$ with the system size
ranged from $16$ to $32$ stepped by $4$ for $p=p_{c}=0.5$, i.e., the
percolating case. While for $p=0.4$ and $p=0.3$, the distribution functions
of $\ln (P_{2})$ with the different system size are also calculated$^{14}$.
A scale invariant form of the IPR distribution is found for the critical
case $p_{c}=0.5$, as well as for the noncritical $p=0.4$ and $p=0.3$. Note
that in the Refs. \cite{5} and \cite{6}, only at the criticality of PRBM
ensemble, a scale invariant form of the distribution of the IPR is
investigated. The shift of the curves implies a scaling of distribution
function of the IPR with respect to $L$. There exist different scaling
exponents $D_{q}$ for the different values of $q$ at $p=0.5$, $0.4$, and $%
0.3 $, as verified by the ensemble averaging $\left\langle
P_{q}\right\rangle $ in ref. \cite{14}. All curves are not strictly
overlapped when we shift them by $\ln \left\langle P_{q}\right\rangle $,
especially for the large $q$. One possible reason is that the sample is not
large enough. Therefore, from Figures 3 and 4 we can conclude that for a
definite $p$ the fluctuations of the IPR have the general statistical
features.

Different from the eigenfunctions of the Anderson metal-insulator
Hamiltonian and PRBM ensemble, the right eigenvectors of Green's-matrix $M$
are the main product of the local field $V$ for the binary composites. In
ref. \cite{14}, we found the scaling of $\left\langle P_{q}\right\rangle $
with respect to the size $L$ for $p_{c}=0.5$, $p=0.4$ and $p=0.3$. By the
relation, $\left\langle P_{q}\right\rangle =L^{-\tau _{q}}$, the scaling
exponent $\tau _{q}$ can be calculated through the slope of $\ln
\left\langle P_{q}\right\rangle $ with $\ln (L)$. The results are shown in
Fig. 5. For $p=0.5$ and $p=0.4,$ we observe the multifractal scaling of $%
\left\langle P_{q}\right\rangle $. Note that for $p=0.4$ it is not a
critical case. While in Ref. \cite{5} and Ref. \cite{6}, the multifractal
scaling is investigated only in the critical ensembles. However, for $p=0.3$%
, only the fractal behavior of the ensemble averaged $\left\langle
P_{q}\right\rangle $ is found and for $q>2$ the extended line of $\tau _{q}$
can be approximately approaching the original point $(0,0)$. By the relation 
$\tau _{q}=D_{q}\times (q-1)$, the values of $D_{q}$ are obtained. Here $%
D_{2}$ is not closely related to the spatial geometry as claimed in Ref. %
\cite{21}.

Finally, to connect with the level spacing statistics, the scaling exponent $%
D_{2}$ and spectral compressibility$^{14}$ $\chi $ are shown in Fig. 6, and
the axial symmetry according to their mean values is found when $p$ is
ranged in the interval $\left[ 0.3,0.5\right] $. The result apparently
disagrees with the claim of Ref. \cite{5}, where $D_{2}$ and $\chi $ satisfy 
$D_{2}+2\chi =1$ for the large $b$, and $2D_{2}+\chi =1$ for the small $b$.
It is also violated with the recent statement$^{21}$, where between $D_{2}$
and $\chi $ the exact relation holds, $\chi =(d-D)/2d$.

In a summary, we have not only presented a detailed study on the local field
distribution near resonance, but also the statistics of local field
distribution by means of the IPR for the disordered binary networks. On the
qualitative level, the main findings are summarized as follows: $1)$ For the
percolating composites, the IPR of the right eigenvectors are illustrated,
as well as the typical local field distributions of the localized and
extended states in the forms of 3D plots. When resonance happens, the fields
are localized within the impurity metallic clusters. For a percolating
network, a large enhancement in the effective linear and nonlinear optical
responses is also found$^{14}$. $2)$ The distribution functions of the IPR $%
P_{q}$ have a scale invariant form in the limit of the large system size $L$
for the critical case, $p_{c}=0.5$, as well as for the noncritical case, $%
p=0.4$ or $p=0.3$. $3)$ The multifractal scaling of the ensemble averaged $%
\left\langle P_{q}\right\rangle $ with the system size $L$ is obtained for $%
p_{c}=0.5$ and $p=0.4$, while for $p=0.3$, only the fractal behavior is
found. $4)$ To relate the eigenvectors correlations to level spacing
statistics, the axial symmetry between $D_{2}$ and $\chi $ is obtained.
Finally, it is worthwhile to emphasize on the extent of the universality of
the IPR distribution at the noncritical case, such as $p=0.4$ or $p=0.3$. In
the previous works$^{5,6}$, only at the criticality, the fluctuations and
scaling of the IPR were investigated.

\begin{figure}[tbp]
\caption{ IPR of right eigenvectors with $q=2$ at the percolating threshold $%
p_{c}$. Here we use a $40\times 40$ sample with $1331$ resonances. }
\label{fig1}
\end{figure}

\begin{figure}[tbp]
\caption{ The typical local field distributions of localized and extended
eigenstates. $a)$ with a high value of IPR; $b$) with a low value of IPR.}
\label{fig2}
\end{figure}

\begin{figure}[tbp]
\caption{ Distribution functions of $\ln (P_{q}/\left\langle
P_{q}\right\rangle )$ with respect to $\ln (P_{q}/\left\langle
P_{q}\right\rangle )$ for $q$ ranged from $2$ to $8$ and for the percolating
networks. Here $L=32$ and the distributions are averaged over $900$ samples. 
}
\label{fig3}
\end{figure}

\begin{figure}[tbp]
\caption{Scale invariant form of distribution functions $\ln (P_{2})$ with
respect to $\ln (P_{2})$ for $p_{c}=0.5.$ }
\label{fig 4}
\end{figure}

\begin{figure}[tbp]
\caption{ Diagram of multifractal scaling $\protect\tau _{q}$ for $p=0.5,0.4$%
, and $0.3$. For $p=0.3$, the extended line of $\protect\tau _{q}$ can be
approaching to the point $(0,0)$ when $q>2$, therefore only fractal behavior
is found. }
\label{fig5}
\end{figure}

\begin{figure}[tbp]
\caption{ The fractal dimension $D_{2}$(circles) and the spectral
compressibility $\protect\chi $(squares) as a function of the parameter $p$. 
}
\label{fig6}
\end{figure}

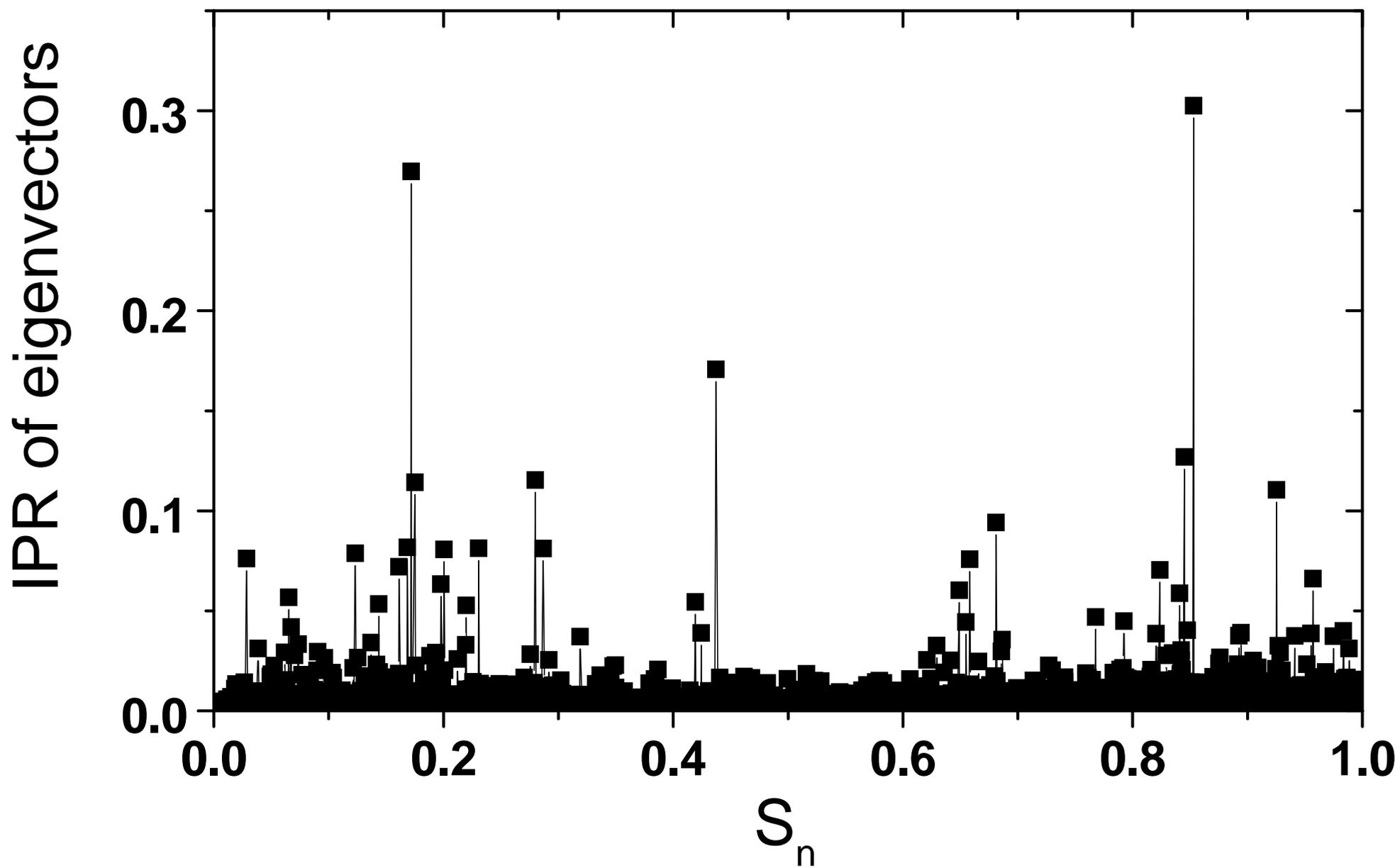
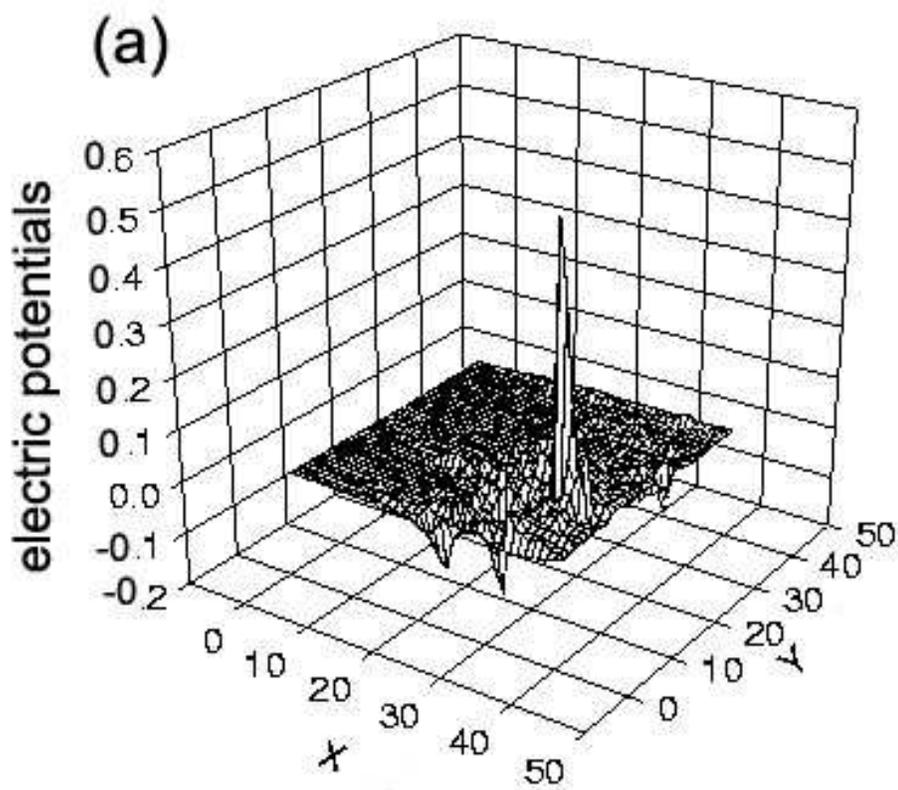
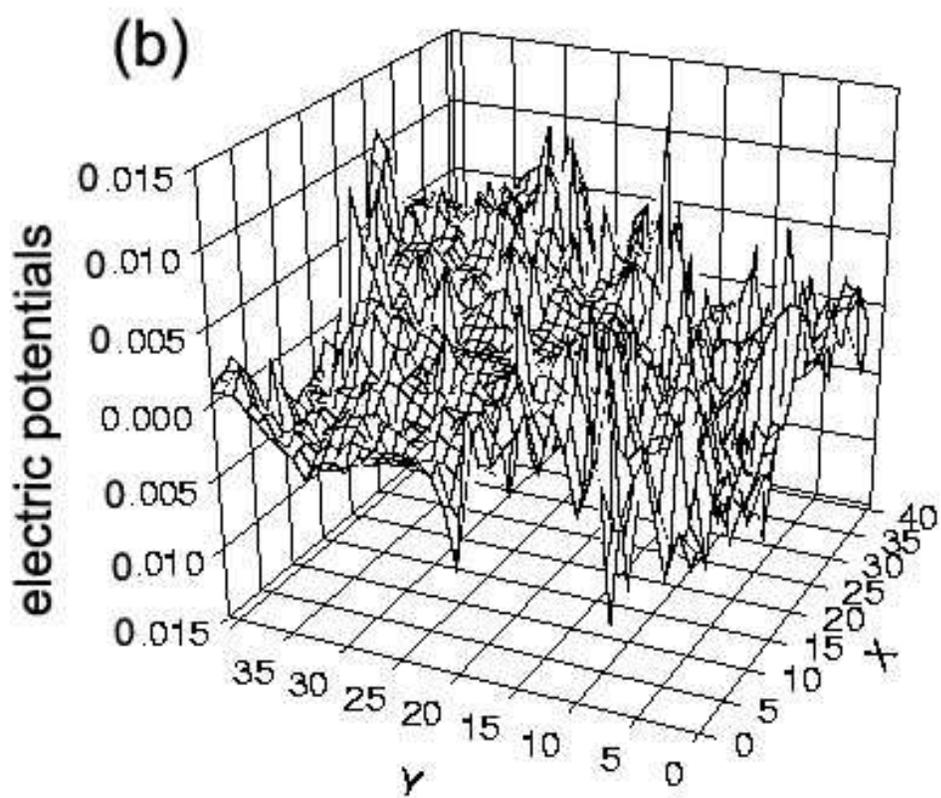
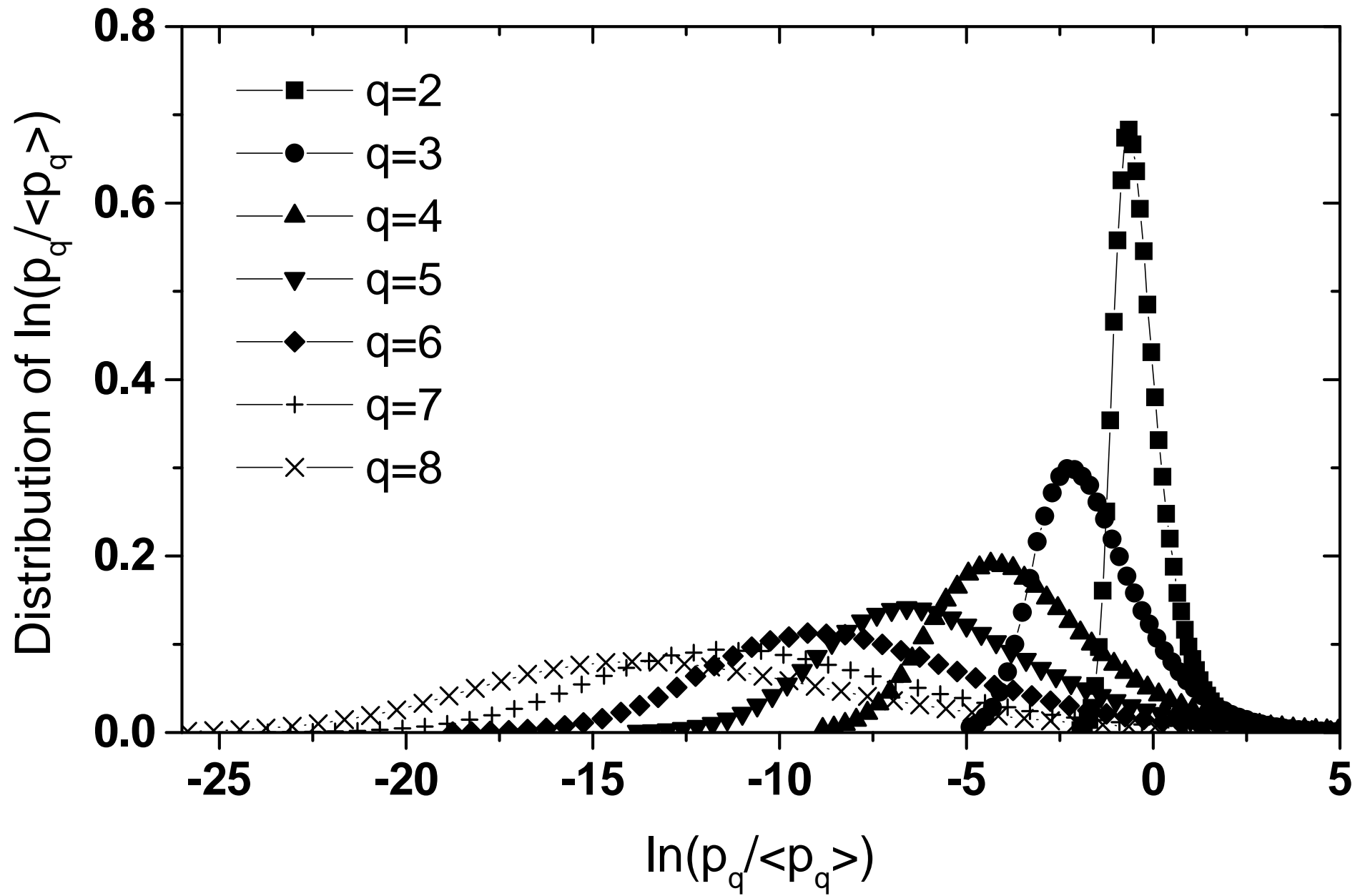
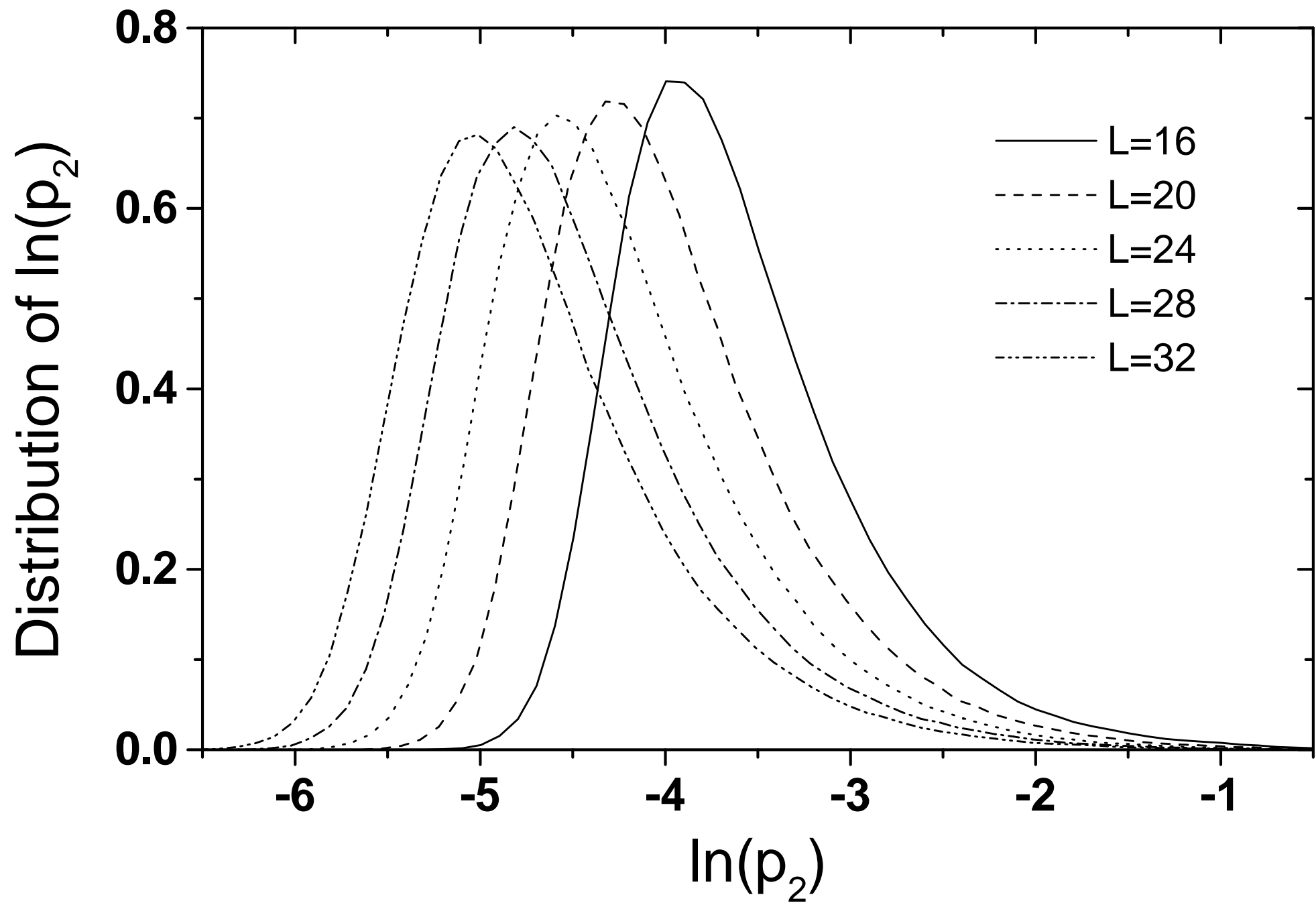
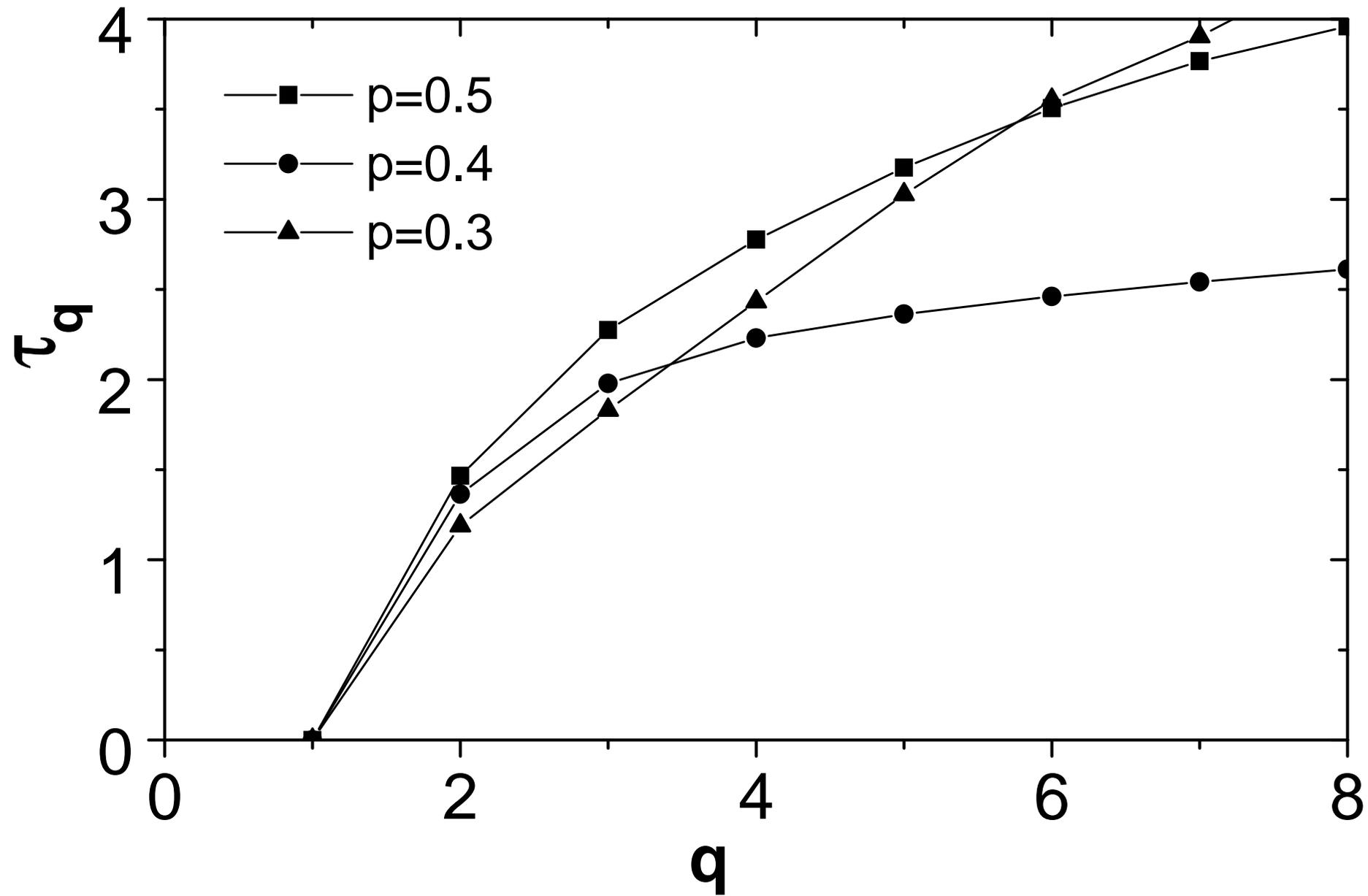
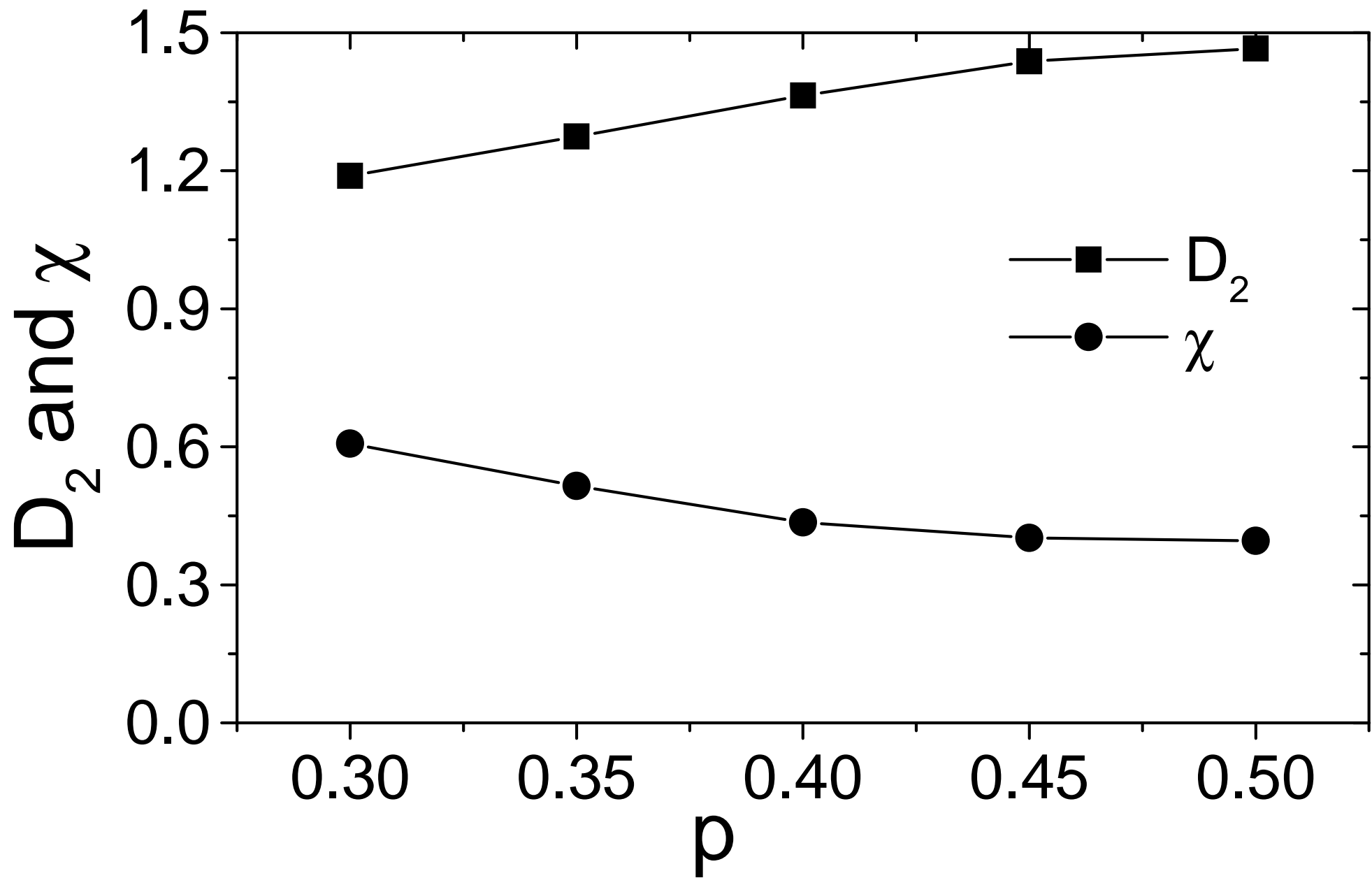
\end{document}